\documentclass[14pt,british]{extarticle}

\newcommand*{\doccode}{ASBHv2}

\usepackage{babel}
\usepackage[utf8]{inputenc}

\usepackage{datetime2}

\DTMsetup{useregional}

\usepackage{fancyhdr}
\usepackage{url}

\usepackage{amsmath}
\usepackage{amsfonts}
\usepackage{circuitikz}
\usepackage{endnotes}
\usepackage{framed}
\usepackage{graphicx}
\usepackage{caption}
\usepackage[version=4]{mhchem}

\usepackage{titlesec}
\titleformat{\section}{\normalsize\bfseries}
   {\thesection}{1em}{}
\titleformat{\subsection}{\normalsize\bfseries}
   {\thesubsection}{1em}{}

\begin{document}


\DTMsavedate{krrevdate}{2026-10-02}

\newcommand*{\revdatefoot}
{\DTMtwodigits{\DTMfetchday{krrevdate}}-\DTMenglishshortmonthname{\DTMfetchmonth{krrevdate}}-\DTMfetchyear{krrevdate}}

\newcommand*{\revdatelong}{\DTMenglishmonthname{\DTMfetchmonth{krrevdate}} \DTMfetchday{krrevdate}, \DTMfetchyear{krrevdate}}

\newcommand*{\revdateyyyymmdd}
{\DTMfetchyear{krrevdate}\DTMtwodigits{\DTMfetchmonth{krrevdate}}\DTMtwodigits{\DTMfetchday{krrevdate}}}


\setlength{\parindent}{0in}
\setlength{\parskip}{0.4in plus0.2in minus0.2in}
\setlength{\voffset}{0in}
\setlength{\topmargin}{0in}
\setlength{\headheight}{0in}
\setlength{\headsep}{0in}
\setlength{\footskip}{0.5in}

\pagestyle{plain}

\noindent

\inputencoding{utf8}

\bibliographystyle{plain}


\long\def\symbolfootnote[#1]#2{\begingroup%
\def\thefootnote{\fnsymbol{footnote}}
     \footnote[#1]{#2}\endgroup}











\begin{center}
{\large\textbf{Analytic Integration of the Lambert W\\
		       Cosmic Fluid Model H(z) Formula}}
 ===============================\\
 \, \\
Ken Roberts\footnote{
	krobe8@uwo.ca or krobe8@gmail.com,
	Formerly Dept of Physics and Astronomy, 
	Western University, London, Canada.
	ORCID:0000-0002-9566-2554},
Francisco Javier Garcia-S\'anchez	\footnote{
	fgarcia@usb.ve,
	Professor Emeritus (in absentia),
	Universidad Simón Bolívar,
	Caracas 1080A, Venezuela.
	ORCID:0000-0002-0868-9792} \\
Subhajit Saha\footnote{
	subhajit1729@gmail.com, 
	Department of Mathematics,
	Panihati Mahavidyalaya, Kolkata 700110,
	West Bengal, India.
	ORCID:0000-0003-0429-6868}	
and 
S. R. Valluri\footnote{
	valluri@uwo.ca or vallurisr@gmail.com,
	Depts of Mathematics, Physics and Astronomy,
	King's University College, 
	Western University, London, Canada.
	ORCID:0000-0001-9211-7609} \\
\revdatelong
\end{center}



\pagestyle{fancy}
\fancyhf{}
\renewcommand{\headrulewidth}{0pt}
\fancyfoot[L]{\doccode-\revdateyyyymmdd}
\fancyfoot[C]{Page \thepage}
\fancyfoot[R]{\revdatefoot}


\textbf{Abstract:}
{The Lambert $W$ (LW) model for the cosmic fluid
	equation of state was proposed by S. Saha and
    K. Bamba in 2019-2020. 
    A recent (early 2026) paper by Dubey, et al, 
    carries out a new fit of the LW model to
    observational data, in order to estimate
    the model's parameters.
    That paper exhibits a formula for the 
    logarithm of the relative Hubble factor at redshift $z$,
    $\ln(H(z)/H_0)$, 
    which is based upon a numerical integration.
    That integral can be evaluated analytically,
    and we present the details in this working paper.
    The resulting analytic expression for $H(z)/H_0$
    may be convenient for exploration of the LW model.
    We present two approximation formulas for
    the Lambert $W$ function, highly accurate for redshifts $z$ between 0 and 3, which may be useful for 
    researchers who do not have ready access to
    a function library with the Lambert $W$ function.
    We also present a simplified proof of
    the analytic formula for another logarithmic integral
    which may be useful for exploring 
    some variants of the LW model.}


\section{Introduction}
\label{sectintro}

In 2019-2020 S. Saha and K. Bamba (SB) suggested
a cosmic equation of state (fluid) model
which rather neatly represents changes
in the expansion rate of the cosmos\cite{SB2020}.
An Equation of State (EoS) of the LW model can
be written as 
\begin{eqnarray}
	\label{eqeos1}
	w_{\text{eff}} = \frac{P}{\rho} &=&
	\Big[
	\theta_1 \ln\big\{W(a)\big\}
	+ 
	\theta_2 \big\{W(a)\big\}^3
	\Big] \, .
\end{eqnarray}
Here $\theta_1$ and $\theta_2$ are
two dimensionless parameters of the model,
$P$ is the pressure of the cosmic fluid,
and $\rho$ is the energy density of the cosmic fluid.
The function $W()$ denotes the Lambert $W$ 
function\cite{CGH1996}.
(The paper \cite{SB2020} has further references.)
Here $a$ is the scale factor of the cosmos,
normalized so that the present day scale factor
is $a_0 = 1$.

\subsection{Scale Factor vs Redshift}

The relative scale factor $a$ and the redshift $z$ are related by
\begin{eqnarray}
	\label{eqaz}
	a \, (1 + z) &=& 1 \, .
\end{eqnarray}
Thus one may transition between formulas which
express the Hubble factor $H$ or some other
property of the cosmos as a function of either $a$ or $z$.
Replace instances of $a$ by $1/(1+z)$,
or replace instances of $z$ by $1/a - 1$.
Taking logarithms in equation (\ref{eqaz}) gives
\begin{eqnarray}
	\label{eqlnaz}
	\ln(a) + \ln(1 + z) &=& 0 \, .
\end{eqnarray}

Taking differentials in equation (\ref{eqlnaz}),
\begin{eqnarray}
	d[\ln(a)] + d[\ln(1+z)] &=& 0 \\
	\frac{da}{a} \, + \, \frac{dz}{1+z} &=& 0
\end{eqnarray}
For integrals of a function $f(a)$ we have
\begin{eqnarray}
	\label{eqintfza}
	\int{\frac{f(a)}{a}} \, da
		+
	\, \int{\frac{f\Big(\frac{1}{1+z}\Big)}{1+z}} \, dz 
	&=& C \, \textrm{\, (arbitrary constant)}   
\end{eqnarray}

We are interested in formulas for the Hubble
factor $H$, considered as a function of $a$ or $z$.
We use subscript 0 to denote present here-and-now;
that is, redshift $z_0 = 0$ and scale factor $a_0 = 1$.
The corresponding present Hubble factor is $H_0$.
We use subscript 1 to denote an arbitrary redshift
$z_1$ or scale factor $a_1$, the two quantities being
related by equation (\ref{eqaz}).\footnote{
In a future working paper,
we will use subscript 2 to denote the redshift $z_2$
and related scale factor $a_2$ at which the
cosmic expansion deceleration parameter $q$
transitions from deceleration to acceleration.}

\subsection{Purpose of this Working Paper}

The Saha and Bamba paper \cite{SB2020} was followed by 
two related papers \cite{BD2020, AS2021}
with interesting further ideas about the model.
The paper \cite{AS2021} reports on a preliminary 
fit of the LW model parameters to observational data.
A quite recent ``fitting" paper was published 
in early 2026 by 
V. C. Dubey, S. Saha and A. Al Mamon\cite{DSAM2026}.
The authors carried out a new analysis of the LW model fit,
using some of the best available data sources.
They obtained $\theta_1 = 0.087 \pm .011$
and $\theta_2 = -3.36 \pm 0.13$.
Their paper \cite{DSAM2026} reports the details.
Their fitted $\theta$ parameters are close
to the original parameter values suggested
by Saha and Bamba in \cite{SB2020}, 
and thereby rekindled our interest in the model.
We wish to explore the LW model further,
and to provide tools to assist others working in the field.

The preprint of \cite{DSAM2026} exhibits an integral
for the natural logarithm of 
the relative Hubble factor $H/H_0$, 
that is $\ln(H/H_0)$, with the
integrand being a function of redshift $z$.
That integral was evaluated using numerical integration.
However, evaluation of the $\ln(H/H_0)$ integral can
be carried out analytically.

An analytic evaluation was first published in
March-2026 by one of us (FJGS) in 
\cite{GS202603,GS202606}, 
and we incorporate portions of his results here.

It is our purpose in this working paper to present
the details for analytical evaluation of the $\ln(H/H_0)$
integral.  That may be of assistance to researchers
who are investigating Lambert $W$ type cosmic fluid models.
In section \ref{sectintegrate} of this working paper, 
we show in detail how to
evaluate the $\ln(H/H_0)$ integral, 
and in section \ref{sectvalidate} we present
a graphical illustration that the analytic
formula agrees with the numerical integration.
As well, we exhibit a Python code segment
which calculates $\ln(H/H_0)$ analytically.

For the convenience of researchers who may not 
have ready access to a function library with the
Lambert $W$ function, in section \ref{sectapprox}
we present two approximation formulas for
the Lambert $W$ function.  One of these is quite
fast, with relative error less than $10^{-4}$ for
redshift values $z$ between 0 and 3 (that is,
scale factor $0.25 < a < 1)$.
The second approximation is also reasonably fast
and has relative error less than $10^{-8}$ 
for redshift values between 0 and 3.

In section \ref{sectjpint} we present a
simplified proof of the analytic formula
for another logarithmic integral
which may be useful for exploring 
some variants of the LW model.
We include that formula here in the
interest of completeness of the tools
provided via this working paper.  
We do not, in this paper, explore variants
of the LW model.

The graphical illustration (figure \ref{fig3p1}
shown later in this working paper) includes a
scatterplot of some observational Hubble data
(OHD), also known as cosmic chronometers (CCH).
We use the list of 32 cosmic chronometers
in a table reported in 2023 by 
Favale, et al, in \cite{FGVM2023}, as that
is the OHD dataset used in \cite{DSAM2026}.
Later papers \cite{TMB2023,FGVM2026} give
more up to date information, and could be
used; our objective here is to have our
calculations be comparable with 
those of \cite{DSAM2026}.
Al Mamon and Saha, in \cite{AS2021}, give a
comprehensive list of references for the
development of cosmic chronometers techniques
(the OHD dataset) over the past two decades.

The OHD data is very helpful in obtaining a
visual, intuitive appreciation of a 
data fit's quality.
The 32 cosmic chronometers of \cite{FGVM2023}
provide good-quality data (tight error bars) 
for redshifts up to about 0.45, and medium-quality
data (some chronometers with tight error bars, 
and some with loose error bars)
for redshifts up to about 1.50.
 
We utilize some techniques described 
by István Mező in his excellent book \cite{MI2022}
about the Lambert $W$ function,
which provides many insightful presentations.

We hope to follow the present working paper
with another working paper which will explore
some possibilities for rewriting the LW model.
We are very impressed by the ``two scenarios"
method used by Saha and Bamba in \cite{SB2020}
to make a rough guess at the $\theta$ parameters.
Results of the latest fit paper \cite{DSAM2026}
suggest that a reduction in parameter count may
be feasible.  

\section{Details of the Analytical Integration}
\label{sectintegrate}

The paper \cite{DSAM2026} uses a formula
which in effect writes the logarithm of 
the relative Hubble factor $H/H_0$ 
as a function of either an arbitrary redshift
$z_1 > 0$ or the corresponding cosmic scale
factor $a_1 = 1/(1+z_1)$ in terms of the
following integrals 
(adapted from eqn (14) of \cite{DSAM2026}).
Here $H_0$ represents the Hubble factor value
(for instance, in km/s/Mpc) 
in our near neighborhood at redshift $z_0 = 0$
or scale factor $a_0 = 1$.
The formula for $H(z)$, equation (15) of \cite{DSAM2026},
is obtained by taking exponentials in the formula for
$\ln(H(z)/H_0)$ and then multiplying by $H_0$.

Hubble factor as a function of redshift $z_1$:
\begin{eqnarray}
	\label{eqlnhz1}
	\ln\Big(\frac{H(z_1)}{H_0}\Big) && \\
	\nonumber
	&=&
	\frac{3}{2} \, \int_{0}^{z_1}{
		\frac{\theta_1 \ln \big(W(\frac{1}{1+z})\big)
			+
			\theta_2 \big(W(\frac{1}{1+z})\big)^3 + 1
		}{1+z} \, dz}
\end{eqnarray}
Hubble factor as a function of scale factor $a_1$:
\begin{eqnarray}
	\label{eqlnha1}
	\ln\Big(\frac{H(a_1)}{H_0}\Big) && \\
	\nonumber
	&=&
	\frac{3}{2} \, \int_{a_1}^{1}{
		\frac{\theta_1 \ln \big(W(a)\big)
			+
			\theta_2 \big(W(a)\big)^3 + 1
	  }{a} \, da}
 \end{eqnarray}
 These two integrals are equal, conversion from
 one expression to the other being achieved by
 interchanging the variables $a_1$ and $z_1$ according
 to equations (\ref{eqaz}) and (\ref{eqintfza}).   
 That is:  Since $a = 1/(1+z)$,
 when $z = z_0 = 0$ 
 the related value of $a$ is $a = a_0 = 1$. 
 When $z = z_1$, the related value
 of $a$ is $a = a_1 = 1/(1+z_1)$.
 Notice that there is a change in the bounds
 of the definite integrals in eqns (\ref{eqlnhz1})
 and (\ref{eqlnha1}), which adjusts for a 
 minus sign required by eqn (\ref{eqintfza}).
 Each integration goes from a lesser coordinate 
 to a greater coordinate: 
 Integration with respect to redshift $z$ goes
 from redshift $z = z_0=0$ at present 
 to redshift $z = z_1$ remotely.
 Integration with respect to scale factor $a$ goes
 from scale factor $a = a_1$ remotely 
 to scale factor $a = a_0 = 1$ at present.

A further simplification of form is obtained by
writing these as logarithmic integrals:
Equations (\ref{eqlnhz1}) and (\ref{eqlnha1}) 
respectively become
\begin{eqnarray}
	\label{eqlnhz2}
	\ln\Big(\frac{H(z_1)}{H_0}\Big)
	&& \\
	\nonumber
	&=&
	\frac{3}{2}\int_0^{z_1}{
		\Big(
		\theta_1 \ln \big(W(\frac{1}{1+z})\big)
		+
		\theta_2 \big(W(\frac{1}{1+z})\big)^3 + 1
		\Big) \, d[\ln(1+z)]} \\
	\label{eqlnha2}
	&=&
	\frac{3}{2}\int_{a_1}^1{
		\Big(
			\theta_1 \ln \big(W(a)\big)
			+
			\theta_2 \big(W(a)\big)^3 + 1
			\Big) \, d[\ln(a)]}
\end{eqnarray}

The analytic integration of equation (\ref{eqlnha2})
was originally obtained by one of us (FJGS) and
published via ResearchGate on 16-March-2026;
see \cite{GS202603} and later update \cite{GS202606}.
The technique used there was somewhat different than
the approach we describe here, but the results
are identical.  We describe the alternative method
for the interest and convenience of the reader.

Considering eqn (\ref{eqlnha2}),
we see that the problem of obtaining an analytic
formula for $\ln(H(z)/H_0)$ reduces to
the problem of analytical evaluations of the
logarithmic integrals of the third power
of Lambert $W$ and of the logarithm of Lambert $W$.
We have used Mező's valuable book \cite{MI2022}, 
pg 21 and thereafter, as our guide.
There are some differences from the 
technique used by Mező since we are 
evaluating logarithmic integrals.
For convenience, we will write out all the details.
The definition of the Lambert $W$ function relates
two variables $x$ and $y$, so that the statement
$y = W(x)$ means precisely that $x = y \, e^y$, 
and vice versa.

We start with
an analytic evaluation of the $\theta_2$ term.
Slightly generalizing the problem, suppose
the task is to determine an indefinite integral
(an antiderivative) of
the logarithmic integral of a positive
power $p > 0$ of Lambert $W$.  
Let $x$ denote the variable of integration.
That is, we wish to evaluate
\begin{eqnarray}
	\label{eqIp}
	I_p(x) &=& \int{\big(W(x)\big)^p}\,d[\ln(x)]
\end{eqnarray}
Let $y = W(x)$, so that $x = y\, e^y$.
Then $\ln(x) = \ln(y) + y$.

Differentiating, we have
\begin{eqnarray}
    d[\ln(x)] = \frac{dx}{x} &=& 
    \frac{dy}{y} + dy = \Big(1 + \frac{1}{y}\Big) \, dy
\end{eqnarray}
The logarithmic integral $I_p(x)$ becomes
\begin{eqnarray}
	I_p(x) &=& \int{y^p }\,\Big(1 + \frac{1}{y}\Big)\,dy \\
	       &=& \int{\big(y^p + y^{p-1}\big)}\,dy
\end{eqnarray}
Since $p > 0$ this is
\begin{eqnarray}
	I_p(x) &=& \int{\big(W(x)\big)^p}\,d[\ln(x)] \\
	&=& \frac{\big(W(x)\big)^{p+1}}{p+1}
	    + \frac{\big(W(x)\big)^p}{p} + C
\end{eqnarray}

Setting $p = 3$ enables us to obtain 
an explicit analytic evaluation of the 
$\theta_2$ term in the formula for $\ln(H/H_0)$.

Now consider the $\theta_1$ term, which
requires a logarithmic indefinite integral (antiderivative)
of the logarithm of $W(x)$.  Let $L(x)$ denote the
desired antiderivative.
We wish to determine
\begin{eqnarray}
	L(x) &=& \int{\Big(\ln\big(W(x)\big)\Big)}\,d[\ln(x)]
\end{eqnarray}
As before, let $y = W(x)$, so that $x = y\, e^y$.
Let $u = \ln(y)$ denote the logarithm of $W(x)$.
Then $u = \ln\big(W(x)\big)$ and 
\begin{eqnarray}
	\ln(x) &=& \ln(y) + y \\
	&=& u + e^u = u + y
\end{eqnarray}
An integral of an integrand 
with respect to a measure which is the 
sum of two measures, is equal to the 
sum of two integrals of that integrand
with respect to each of the measures.
Thus we have
\begin{eqnarray}
	L(x) &=& \int{\Big(\ln\big(W(x)\big)\Big)}\,d[\ln(x)] \\
	&=& \int{\Big(\ln\big(W(x)\big)\Big)}\,\big(d[u]+d[y]\big) \\
	&=& \int{u}\,d[u]
	  + \int{u}\,d[y] \\
	&=& \int{u}\,d[u]
	+ \int{\ln(y)}\,d[y] \\
	&=& \frac{u^2}{2} + y \ln(y) - y + C \\
	&=&
	\frac{\Big(\ln(W(x))\Big)^2}{2} 
	+ W(x)\ln(W(x)) - W(x) + C
\end{eqnarray}

Finally, we need the logarithmic indefinite integral (antiderivative) of the constant 1, call it $K(x)$.  
It is
\begin{eqnarray}
	K(x) &=& \int{1}\,d[\ln(x)] \\
	 &=& \int{\frac{1}{x}}\,dx \, = \, \ln(x) + C%
\end{eqnarray}

Putting those three indefinite integrals
$I_p(x)$, $L(x)$ and $K(x)$ together
allows us to obtain an analytic expression for the
antiderivative which appears in the $\ln(H/H_0)$ formula.
From equation (\ref{eqlnha2}), and replacing $x$
by $a$, we get:
\begin{eqnarray}
	\label{eqlnhz5}
	\ln\Big(\frac{H(z_1)}{H_0}\Big)
	&=& \\
	\nonumber
	&=&
	\frac{3}{2}\int_{a_1}^1{
	\Big(
	\theta_1 \ln \big(W(a)\big)
	+
	\theta_2 \big(W(a)\big)^3 + 1
	\Big) \, d[\ln(a)]} \\
	&=&
	\label{eqLI3K}
	\frac{3}{2} 
	\Big[\theta_1 L(a) 
	   + \theta_2 I_3(a) + K(a)\Big]_{a_1}^{1}	
\end{eqnarray}
For further clarity, we can
write $y = W(a)$, evaluate at $a_1 = a(z_1)$ 
and at $a_0 = 1$, and simplify the expressions.
At $z_0=0$ we have $a_0=1$ and $W(a_0) \approx 0.567$.
Mező uses the symbol $\Omega$ for $W(1)$ and we
follow his notation.  
We have
\begin{eqnarray}
	\ln(W(1)) = \ln(\Omega) = -\Omega
\end{eqnarray}
which is an important property of the
$\Omega$ value, facilitating transfers
of ``structure" between the logarithmic
and linear mathematical worlds. 	
Thus the symbol $\Omega$ shows up in the 
following definite integrals in two roles,
as the value of $W(1) = \Omega$
and in the value of $\ln(W(1)) = -\Omega$.
There is some cancellation among the $\Omega$ values upon simplification.

For convenience, we will write out separately each of
the three evaluated definite integrals which 
appear in eqn (\ref{eqLI3K}):

The $L(a)$ integration gives
\begin{eqnarray}	
	\label{eqL}
	\frac{3}{2} \Big[ \theta_1 \, 
	L(a) 
	\Big]_{a_1}^{1}
    &=& \\
    \nonumber	
    \frac{3}{2} \, \theta_1 	
    &\Big(&
	\frac{\Big(\ln(W(1))\Big)^2}{2} 
	+ W(1)\ln(W(1)) - W(1) \Big) \\	
	- \frac{3}{2} \, \theta_1 
	&\Big(&
	\nonumber
	\frac{\Big(\ln(W(a_1))\Big)^2}{2} 
	+ W(a_1)\ln(W(a_1)) - W(a_1)
	\Big)
\end{eqnarray}
and the first (constant) term in brackets
simplifies to $-\Omega^2/2 - \Omega$.

The $I_3(a)$ integration gives
\begin{eqnarray}	
	\label{eqI3}
	\frac{3}{2} \Big[ \theta_2 \, 
	I_3(a) 
	\Big]_{a_1}^{1}
	&=& \\
	\nonumber	
	\frac{3}{2} \, \theta_2 	
	&\Big(&
    	\frac{\big((W(1))\big)^4}{4} 
	  + \frac{\big((W(1))\big)^3}{3}
	 \Big) \\	
	- \frac{3}{2} \, \theta_2 
	&\Big(&
	\nonumber
    	\frac{\big((W(a_1))\big)^4}{4} 
	  + \frac{\big((W(a_1))\big)^3}{3}
	\Big)
\end{eqnarray}
and the first (constant) term in brackets
simplifies to $\Omega^4/4 + \Omega^3/3$.
    
The $K(a)$ integration gives
\begin{eqnarray}	
    \label{eqK}
    \frac{3}{2} 
    \Big[ K(a)\Big]_{a_1}^{1}
    &=&
    - \, \frac{3}{2} \, \ln(a_1)
\end{eqnarray}

The analytic formula for $\ln(H(z)/H_0)$
is the sum of the formulas in eqns
(\ref{eqL}), (\ref{eqI3}) and (\ref{eqK}).
The analytic formula for $H(z)/H_0$ is
the exponential of that.
We can substitute $1/(1+z)$ for $a$ in the formula.

In section \ref{sectjpint} below we use similar
methods to calculate an analytic evaluation of
another integral $J_p(x)$, which is the $I_p(x)$
integral with an additional $\ln(W(x))$ factor in
the integrand.  That may be useful for variants
of the LW model.  

\section{Verification of the Analytic Formula}
\label{sectvalidate}

\begin{figure}
	\centering
	\includegraphics{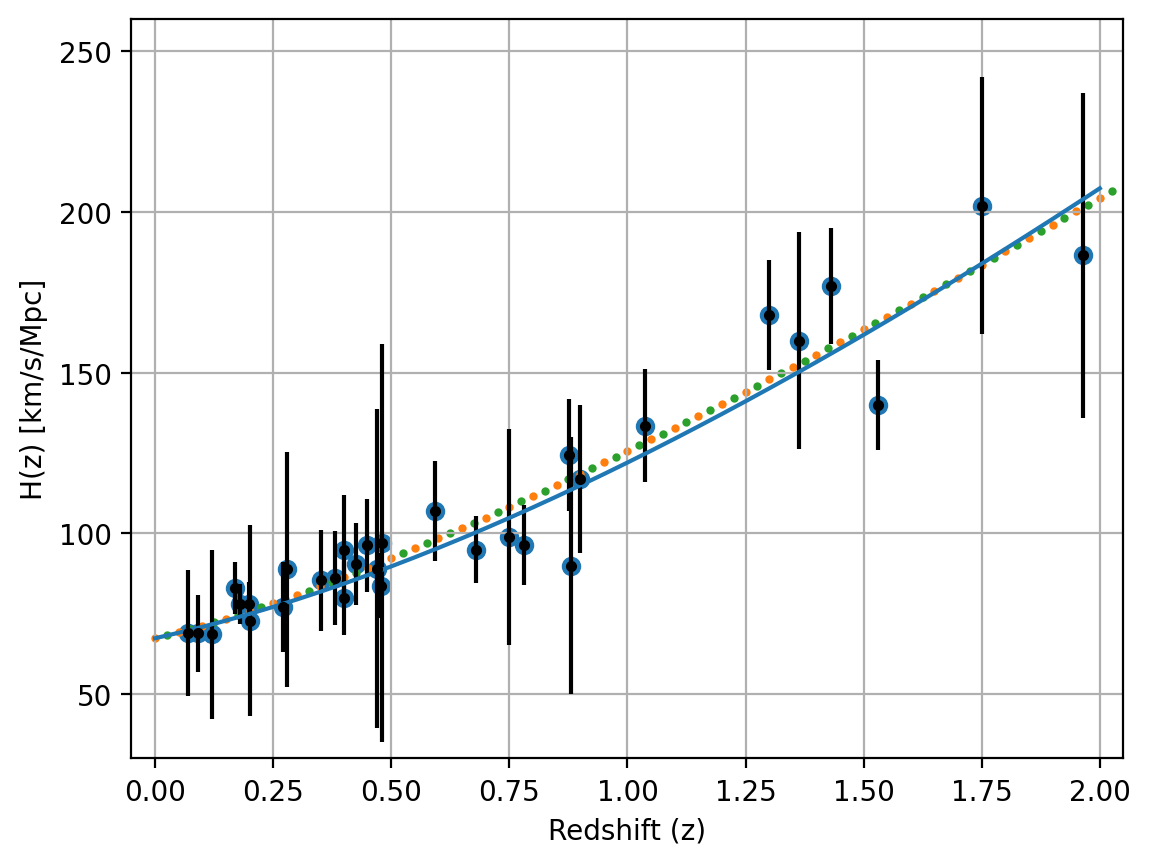}
	\caption{H(z) as a function of redshift $z$. 
	Dark blue dots are $H(z)$ values for the OHD cosmic 
	chronometers, with error bars, from \cite{FGVM2023}. 
	Orange dots are H(z) calculated by formula in 
	\cite{DSAM2026}, using numerical integration,
	with parameters $\theta_1=0.087$ and $\theta_2=-3.36$.
	Green dots are $H(z)$ calculated by the analytic
	formula described in this paper.  The solid
	blue line is a $\Lambda$CDM model based upon
	$\Omega_m=0.3257$.}
	\label{fig3p1}
\end{figure}

The accuracy of the analytic formula can be
verified by sample calculations.  However a
visual verification may be more convincing.
In Figure \ref{fig3p1} we show the observational Hubble
data (OHD, ``cosmic chronometers") described
in \cite{FGVM2023} as dark blue dots
with error bars, and also show three curves of 
$H(z)$ vs $z$.  This figure can be compared
with figure 3 of the paper \cite{DSAM2026},
which was produced by numerical integration of
equation (\ref{eqlnhz1}).
In our figure the orange dots correspond
with every other one of the pink dots in 
figure 3 of \cite{DSAM2026}, and were 
produced by independently calculated
numerical integration using the $\theta$ parameters
determined in that paper to be the best fit.
That is, $\theta_1 = 0.087$ and $\theta_2 = -3.36$.
The green dots in our figure were instead
produced by the analytic formula which we
described above, for $z$ values in between 
the $z$ values of the orange dots, using the
same $\theta$ parameters as for the orange curve.
For comparison, the blue curve in our figure
was calculated via the ${\Lambda}$CDM model,
as described in equation (16) of the paper
\cite{DSAM2026}, and using an estimated value
of 0.3257 for $\Omega_m$, the present density
of dark matter.  We believe that the discrepancy
between the ${\Lambda}$CDM curve in figure 3 of 
\cite{DSAM2026} and our figure is because the
two curves used different values of $\Omega_m$.

Appendix A has the Python code used 
to calculate $H(z_1)/H_0$ given the $\theta$
parameters.

\section{Approximation of 
	     Lambert $W(1/(1+z))$ Function for $z < 3$}
\label{sectapprox}

Suppose that one's choice of computer software 
does not provide a routine to calculate the Lambert $W$ function.  The logarithm of the normalized Hubble function
in the LW model, expressed in terms of the redshift $z$
(see equation (4) in \cite{GS202606}), is:
\begin{eqnarray}
	\label{eqHz4}
	\ln\Big(\frac{H(z)}{H_0}\Big) &=&
	\frac{3}{2}\Big\{[\theta_1\big[
	\frac{y^2}{2} + y - \frac{\ln(y^2)}{2}+\alpha
	\big] \\
	&+&
	\nonumber
	\,\,\,\,\,\,\,\,\,\, \theta_2\big[
	\beta - \frac{y^4}{4} - \frac{y^3}{3}
	\big]
	\Big\}
\end{eqnarray}
Here the symbol $y$ represents a function
of the redshift $z$, namely
\begin{eqnarray}
	\label{eqy1z}
	y &=& W\Big(\frac{1}{1+z}\Big) \, .
\end{eqnarray}
The symbols $\theta_1$ and $\theta_2$ are the
fitting parameters of the LW model.
The symbols $\alpha$ and $\beta$
represent two constants that arise
from taking into account the
starting point, $z = 0$,
for calculation of
the definite integral (\ref{eqlnhz2}).
These two constants  are given by:
\begin{eqnarray}
	\alpha &=& - \Big[W(1)+\frac{W(1)^2}{2}\Big]
		\approx -0.7279690463 \\
	\beta &=& \Big[\frac{W(1)^3}{3}+\frac{W(1)^4}{4}\Big]
		\approx 0.08667242270
\end{eqnarray}

The RHS of (\ref{eqHz4}) contains first- to fourth-order powers of $W(1/(1+z))$, first and second-order powers of $\ln(W(1/(1+z)))$, and 
constants.  
Note, incidentally, that we are using the
principal branch (the $W_0$ branch) of the Lambert $W$
function, since the argument $1/(1+z)$ is a positive real
number and the desired result is also a real
number.

\subsection{Why use an Approximation Formula?}

The calculation of the formula (\ref{eqHz4})
may involve more computer work than is necessary,
if one does not have an efficient implementation of
the Lambert $W$ function.
Even if one does have an efficient implementation
of Lambert $W$, the range of values of $z$ used
in a typical fitting computation is constrained.
For practical computational reasons, it can be
beneficial to express $W(1/(1+z))$ in equation
(\ref{eqy1z}) as 
an approximate expression containing only elementary
functions.  That would circumvent the need to
calculate $y = W(a) = W(1/(1+z))$ by inverting
$y \, e^y = a$, its transcendental 
implicit defining equation,
through numerical iterative algorithms, typically
of the Halley's method type.

\subsection{Considerations and conditions}

To be effective, an analytical approximation should
balance accuracy and simplicity to ensure ease of
calculation.  Such a balance is easier to achieve
when the domain and range of the function to be
approximated are restricted to relatively small
intervals.  
Since theoretical values of redshift $z$ lie in the 
interval $z \in [0,\infty]$, the values of the
argument of $W(a)$, the cosmic scale factor 
$a = 1/(1+z)$ theoretically fall within the interval
$a \in [0,1]$.
Nevertheless, in many cases we are interested in 
the normalized Hubble function (\ref{eqHz4})
for moderate redshift values, that is, those that lie approximately within the interval $z \in [0,3]$.
Thus, in such cases, the argument of $W(a)$ is confined
to an even smaller interval of $a \in [1/4,1]$,
which means that good analytical approximations of $W(a)$
are possible.
Analytic approximations of the $W$ function's
principal branch have been reported in the literature
for small values of its argument.  However, we propose
below two approximations specifically tailored for the
interval of interest.

\subsection{Proposed approximations}

We express the following approximation formulas in terms
of the redshift $z$.  The corresponding formula 
expressed in terms of the cosmic scale factor $a$
can be obtained by writing $(1/a)-1$
in place of $z$.

\begin{figure}
	\centering
	\includegraphics{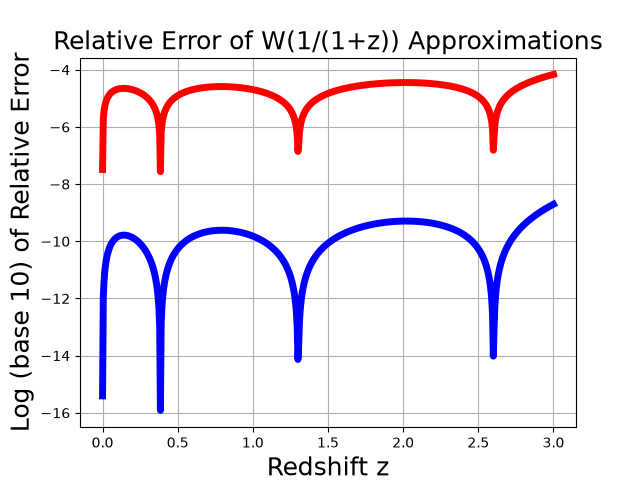}
	\caption{Relative error of two approximations
		to the Lambert $W(1/(1+z))$
		function, evaluating $W(a)$ or
		$W(1/(1+z)$ as described in section 
		\ref{sectapprox}, for redshift $z$ in $0 < z < 3$.
		Red curve (upper graph) shows approximation A,
		with relative error less than $10^{-4}$.
		Blue curve (lower graph) shows approximation B, 
		with relative error less than $10^{-8}$.}
	\label{fig4p1}
\end{figure}

The following two analytic approximations of $W(1/(1+z))$
can be used, depending on the required accuracy.
Note that the relative error graphs in Figure \ref{fig4p1}
are the relative error in the Lambert $W$ approximation
formula itself.
They are not errors in
the LW model's formula for $w_{eff}$ at redshift $z$ as 
determined by equation (\ref{eqeos1}).
Those error calculations would depend upon choices
of the two $\theta$ parameters of the model.

{\bf{First Approximation (A):}}
For rough work one can use 
a very simple approximation comprised of a single
logarithm with only three parameters in its argument.
Within the interval of interest $z \in [0,3]$ its 
{\bf{relative error}} is less than $10^{-4}$ as shown
by the red curve in figure \ref{fig4p1}.  This figure
shows the base-10 logarithm of the relative error data.
\begin{eqnarray}
	\label{eqy1}
    W\Big(\frac{1}{1+z}\Big) &\approx& y_1 =
    \frac{1}{2} \ln\Big(
    \frac{b_1 z + 2}{b_2 z + b_3}\Big)
\end{eqnarray}
where the approximation parameters are
$b_1 = 0.689837, b_2 = 0.687788$, and $b_3 = 0.643303$.

{\bf{Second Approximation (B):}}
For more exact work, one can use 
a second approximation, derived from the former
approximation $y_1$ by a single step of Newton's method.
It includes a second evaluation of a logarithm.
Within the interval of interest
$z \in [0,3]$, its
{\bf{relative error}} is less than $10^{-8}$ as shown
by the blue curve in figure \ref{fig4p1}.
\begin{eqnarray}
	\label{eqy2}
	W\Big(\frac{1}{1+z}\Big) &\approx& y_2 =
	y_1 \Big(\frac{1-\ln\big(y_1(1+z)\big)}{1+y_1}\Big)
\end{eqnarray}
The approximation parameters for this formula are
the same as before in formula A, namely 
$b_1 = 0.689837, b_2 = 0.687788$, and $b_3 = 0.643303$,
since the Newton's method iteration provides the
desired reduction in relative error from the $y_1$
approximation.

For convenience, we record here the reasoning
which led to the above formula (\ref{eqy2}) for
a Newton's method refinement of formula (\ref{eqy1}).
Let $y_0 = W(a)$ be the correct value of 
$W(a) = W(1/(1+z))$ and apply the implicit definition
of Lambert $W$ to get $a = y_0\,e^{y_0}$.
Take natural logarithms of both sides to get:
\begin{eqnarray}
	\ln(a) &=& \ln(y_0) + y_0
\end{eqnarray}
Define the objective function $f$ to be used in the
iteration of Newton's method, as a function of
variable $y$, in place of exact value $y_0$:
\begin{eqnarray}
	f(y) &=& \ln(y) + y - \ln(a)
\end{eqnarray}
Calculate its first derivative with respect to $y$,
which is:
\begin{eqnarray}
	f^{'}(y) &=& \frac{1}{y} + 1
\end{eqnarray}
One iteration of Newton's method, starting from an
initial guess of $y_1 \approx y_0$,
yields the expression for
the next guess $y_2 \approx y_0$:
\begin{eqnarray}
	y_2 &=& y_1 - \frac{f(y_1)}{f^{'}(y_1)} \\
	&=& y_1 
	- \frac{\ln(y_1) + y_1 - \ln(a)}{\frac{1}{y_1} + 1} \\
	&=& y_1 \Big( 
	1 - \frac{\ln(y_1) + y_1 - \ln(a)}{1 + y_1}\Big) \\
	&=& y_1 \Big( 
	\frac{1 - \ln(y_1) + \ln(a)}{1 + y_1}\Big)
\end{eqnarray}
Substituting $1/(1+z)$ for $a$ in the $\ln(a)$ term
and simplifying gives:
\begin{eqnarray}
	y_2 &=&  y_1 \Big( 
	\frac{1 - \ln\big(y_1(1+z)\big)}{1 + y_1}\Big)
\end{eqnarray}

\section{Evaluation of a Related Integral}
\label{sectjpint}

A recent paper \cite{JA20260812} by another
author has, extending our methods,
performed the analytic evaluation of 
a related indefinite logarithmic integral
\begin{eqnarray}
	\label{eqJp}
	J_p(x) &=& 
	\int{\big(W(x)\big)^p}\,\ln(W(x))\,d[\ln(x)]
\end{eqnarray}
where $p$ is a non-negative integer.
This is the $I_p(x)$ integral of eqn (\ref{eqIp})
with an additional factor $\ln(W(x))$ in
the integrand.

To clarify their result and to provide a more
comprehensive toolkit for exploring the LW model, 
we work through a detailed evaluation 
of the $J_p(x)$ integral.
Consider any real $p > 0$.
Using the notational conventions
from section \ref{sectintegrate} above
let $y = W(x)$.
As before, we have
\begin{eqnarray}
	x &=& y \, e^y \\
	\ln(x) &=& \ln(y) + y \\
	d\ln(x) &=& d\ln(y) + dy \\
	&=&
	    (1 + \frac{1}{y}) \, dy
\end{eqnarray}

We start with another indefinite integral expression,
$M_q(y)$, with an integrand which is similar to the
integrand of $J_p(x)$ in structure, 
but which is an ordinary (not logarithmic) integral,
with the free variable being $y$ not $x$ and the power
appearing in the integrand expression being $q$ not $p$.
That is, 
\begin{eqnarray}
	M_q(y) &=& 
	\int{y^q \, \ln(y) } \,dy 
\end{eqnarray}
Using integration by parts, integrating $y^q$ and
differentiating $\ln(y)$, the $M_q(y)$
integral evaluates as
\begin{eqnarray}
	M_q(y) &=& 
	\frac{1}{q+1} \, y^{q+1} \ln(y)
	- \frac{1}{q+1} \int{y^{q+1} \, \frac{1}{y}} \,dy \\
	&=& \frac{1}{q+1} \, y^{q+1} \ln(y)
	- \frac{1}{(q+1)^2} \, y^{q+1} + C
\end{eqnarray}
The $M_q(y)$ indefinite integral is valid for all $q > -1$.

Now let's evaluate the $J_p(x)$ integral.
We allow any real $p > 0$.
With the above substitution of $y = W(x)$,
the $J_p(x)$ indefinite integral becomes
\begin{eqnarray}
	J_p(x) &=& 
	\int{y^p \, \ln(y) } \, (1 + \frac{1}{y}) dy \\
	&=& \int{y^p \, \ln(y) } \, dy 
	\, + \, \int{y^{p-1} \, \ln(y) } dy \\
	&=& M_p(y) + M_{p-1}(y)
\end{eqnarray}

Hence we obtain
\begin{eqnarray}
	J_p(x) &=& \frac{1}{p+1} \, y^{p+1} \ln(y)
	- \frac{1}{(p+1)^2} \, y^{p+1} \\
	&+&
	\frac{1}{p} \, y^{p} \ln(y)
	- \frac{1}{p^2} \, y^{p} + C
\end{eqnarray}
where $y$ denotes $W(x)$.

\section{Conclusion}

In this working paper we presented
the details for analytical evaluation of the $\ln(H/H_0)$
integral which appears in the Lambert $W$ (LW) model
described by Saha, Bamba, and colleagues in
references \cite{SB2020} to \cite{DSAM2026}.
We presented a graphical validation of the
resulting formula for $H(z)$.
As further useful tools for investigating
the LW model, we presented approximation formulas
for evaluating Lambert $W(1/(1+z))$ appearances in the model
which are very accurate (relative error less
than $10^{-4}$ or $10^{-8}$ depending upon
the formula chosen) for redshifts $z$ in $0 < z < 3$,
or corresponding scale factor $a$ in $0.25 < a < 1$.

It is our hope that the techniques described in
this working paper will be useful for researchers
who are studying properties of the Lambert $W$ 
cosmic fluid model.

\newpage
\begin{appendix}
	{\bf{Appendix A:}}
	
For convenience, we provide the Python code used 
to calculate $H(z_1)/H_0$ given the $\theta$
parameters.  Note that the lambertw function
must be imported from scipy.special.

\begin{verbatim}
# Fcn to calculate H(z)/H0 using the analytic formula.
from scipy.special import lambertw
def relHzf(z1) :
	# The argument z1 is the value of redshift z.
	# Convert redshift arg to scale factor arg a1.
	a1 = 1/(1+z1)
	Omega=lambertw(1).real
	lnOmega = -Omega
	# Formula for L term of log(H/H0).
	y = lambertw(a1).real
	lny = log(y)
	Lterma0 = (lnOmega**2)/2 + Omega*lnOmega - Omega
	Lterma1 = (lny**2)/2 + y*lny - y
	Lterm = (3/2)*theta1*(Lterma0 - Lterma1)
	# Formula for I3 term.
	I3terma0 = (Omega**4)/4 + (Omega**3)/3
	I3terma1 = (y**4)/4 + (y**3)/3
	I3term = (3/2)*theta2*(I3terma0 - I3terma1)
	# Formula for K term.
	Kterma0 = 0
	Kterma1 = log(a1)
	Kterm = (3/2)*(Kterma0 - Kterma1)
	lnrelHz = Lterm + I3term + Kterm
	return exp(lnrelHz)
\end{verbatim}
	
\end{appendix}
	
[end]

\end{document}